\documentstyle[prl,aps,preprint]{revtex}
\begin {document}
%\twocolumn[

\draft
\title{Entropic droplets and activated events near the glass transition 
of a random heteropolymer}
\author{Shoji Takada and Peter G. Wolynes}
\address{School of Chemical Sciences, University of Illinois, Urbana, IL, 61801}
\date{\today}
\maketitle

\widetext
\begin{abstract}
The barriers between metastable states near 
the glass transition of a random heteropolymer are studied 
using replicas by describing inhomogeneous states.
The instanton solution for a replica space free energy functional 
is found numerically to estimate the size of activation barriers 
and of the critical nuclei themselves between the dynamic 
and the static glass transition temperatures.
\end{abstract}

%]
%\narrowtext

Quantifying escape from configurational traps on a rugged free energy landscape 
is important for understanding the dynamics of spin glasses\cite{SGdyna}, 
structural glasses\cite{Angell95}, and folded proteins\cite{Frauenfelder91}, 
and for protein folding\cite{Bryngelson95}.
A first step towards understanding barriers is to appreciate the organization of the 
stable minima\cite{Mezard87}. This has only been carried out completely for 
infinite range spin glasses. That organizational structure 
inspires many of the dynamical theories
\cite{Bryngelson95,Garel88,Shakhnovich89,Sasai90,Takada96}.
Barriers in mean field spin glasses scale with the system size, but
the finite range of interactions allows escape from 
traps through localized reconfigurations with finite barriers in the thermodynamic 
limit. The Vogel-Fulcher law for viscosity of structural glasses has been 
explained through such a mechanism\cite{Kirkpatrick87}. 
Various mean field theories of structural glasses resemble those for 
spin glasses lacking reflection symmetry. 
In strict mean field theory these models undergo a dynamical transition at 
a temperature $T_{\rm A}$, where a macroscopic number of frozen free 
energy minima appear, and a static transition at $T_{\rm K}$, 
the Kauzmann temperature, where the configurational entropy of the minima disappears.
Kirkpatrick and Wolynes (KW) pointed out that that individual free energy minima 
between $T_{\rm A}$ and $T_{\rm K}$ will be inherently unstable for 
short range interaction models because of entropic droplets. 
The extensive configurational entropy provides a driving force for a 
localized region in a local minimum to reconfigure and escape the trap. 
Their analysis gave a modified Vogel-Fulcher law while a later 
scaling picture incorporating entropic droplets gave precisely the 
usual form used empirically\cite{Kirkpatrick89}.
Parisi has presented a novel instanton 
argument in replica space yielding the original KW form\cite{Parisi94}.
Here, we use replica instanton calculations to quantify reconfiguration 
barriers for the random heteropolymer. 

Reconfiguration barriers determine the 
configurational diffusion coefficients that enter the theory of 
protein folding times\cite{Bryngelson95,Socci96}. 
At low $T$, 
trap escape is rate limiting. 
Whether the escape barrier is extensive can be tested experimentally 
in a crude fashion since proteins are mesoscopic 
with finite chain length $N$. 
The size scaling is presently controversial\cite{Socci96,Gutin96,Thirumalai95}. 
The barriers computed in the strict mean field limit of 
uniform transitions are of order
$\sim0.1Nk_BT_{\rm K}$ at the static transition $T_{\rm K}$ using parameters 
fit to the lattice model thermodynamics\cite{Takada96}. 
This seems to be consistent with recent simulation results at low temperature\cite{Gutin96}. 
Our calculations suggest entropic droplets are often large in the random heteropolymer 
so that the mean field arguments are a good starting point for mesoscopic 
systems of the size of the smaller proteins. 
One caveat is that polymers have additional entanglement 
constraints not present in structural or spin glasses. 
These are neglected here.
 
We focus on the temperature range near the glass-like transition 
of a finite-size heteropolymer in a poor solvent so that it is collapsed. 
The droplet we deal with is a small globular region of the polymer 
that may take on many configurations, 
local in space but not necessarily in sequence, buried in a remaining frozen glassy 
portion.
We utilize replica formalism and derive a Landau-like free 
energy functional in terms of the Debye-Waller factor for a residue which plays
the role of an Edwards-Anderson order parameter which is 
taken as spatially varying.

We introduce the standard bead contact Hamiltonian for a random heteropolymer 
which includes finite range random interactions between monomers;
%\twocolumn[
%\widetext

\begin{eqnarray}
H&=& k_BT \sum_{i} {({\bf r}_{i+1}-{\bf r}_{i})^2\over 2a^2}
     +{1\over 2} \sum_{i\neq j}b_{ij} {\rm e}^{ -({\bf r}_{i}-{\bf r}_{j})^2/\sigma^2}
     +{c\over 6} \sum_{i\neq j\neq k} {\rm e}^{- \left[ ({\bf r}_{i}-{\bf r}_{j})^2
                                      + ({\bf r}_{j}-{\bf r}_{k})^2\right]/\sigma^2},
\end{eqnarray}
where ${\bf r}_i$ represents positions of Kuhn segments ($i=1\sim N$), 
$a$ is the Kuhn length, $b_{ij}$ and $c$ 
are the second and third virial coefficients, respectively, and $\sigma$ is the characteristic 
length of interactions. 
$b_{ij}$ are Gaussian random variables with distribution, 
\protect{$P\left( b_{ij}\right) = (2\pi b^2)^{-1/2} \exp \left[ -(b_{ij}-b_0)^2/ 2b^2 \right]$}.

The free energy $F_{\rm av}$ is first averaged over this random interaction using
standard replica formalism\cite{Mezard87}, 
to obtain an effective Hamiltonian, $H_{\rm eff}=H_0+H_1+H_2$ where 
$H_0=k_BT\sum_{i\alpha}{({\bf r}_{i+1}^\alpha-{\bf r}_{i}^\alpha)^2/ 2a^2}$ represents 
the elasticity, 
$H_1=1/2(b_0-\beta b^2/2)\sum_{i\ne j\alpha}
\exp [-{({\bf r}_{i}^\alpha-{\bf r}_{j}^\alpha)^2/ \sigma^2}]
+c/6\sum_{i\ne j\ne k\alpha}{\rm e}^{- \left[ ({\bf r}_{i}^\alpha-{\bf r}_{j}^\alpha)^2
        + ({\bf r}_{j}^\alpha-{\bf r}_{k}^\alpha)^2\right]/\sigma^2}$ are effective 
homopolymeric interactions and $n$ is the number of replicas. 
The inter-replica interaction part $H_2$, responsible for breaking ergodicity, 
can be written in terms of 
an overlap order parameter function $Q_{\alpha\beta}$ defined by
$Q_{\alpha\beta}({\bf r}_1,{\bf r}_2)
=\sum_i \delta ({\bf r}_1-{\bf r}_i^\alpha) 
        \delta ({\bf r}_2-{\bf r}_i^\beta)$\cite{Shakhnovich89},
where $\alpha$ and $\beta$ are replica indices;
\protect{
\begin{eqnarray}\label{eq:H2}
H_2 &=&-{\beta b^2\over 4} \int {\rm d}{\bf r}_1{\rm d}{\bf r}_2
                              {\rm d}{\bf r}_3{\rm d}{\bf r}_4
  \left[ \sum_{\alpha \neq\beta} Q_{\alpha\beta}({\bf r}_1,{\bf r}_2)
                                    Q_{\alpha\beta}({\bf r}_3,{\bf r}_4) \right]
   {\rm e}^{- \left[ ({\bf r}_1-{\bf r}_3)^2
                                      + ({\bf r}_2-{\bf r}_4)^2\right]/\sigma^2}.
\end{eqnarray}
}

Exact integration over the bead variables is difficult, but 
a variational approach extended into replica 
space\cite{Edwards88,Sasai90,Takada96} can be used 
with a reference Hamiltonian $H_{\rm ref}$ to calculate a variational free 
energy $F_{\rm var}$ by finding extrema of
$F_{\rm var}\equiv -k_BT\ln Z_{\rm ref}+\langle H_{\rm eff}-H_{\rm ref}\rangle 
$. $Z_{\rm ref}$ is the partition function for $H_{\rm ref}$ and
$\langle \cdots \rangle$ means the average with $H_{\rm ref}$.
The physical free energy $F_{\rm av}$ is 
$\lim_{n\rightarrow 0}F_{\rm var}^*/n$, where $*$ means its extreme value.
We use the same reference Hamiltonian as ref.I (with $C=0$ in ref.I),
\begin{eqnarray}\label{eq:refH}
\beta H_{\rm ref}&=&\sum_{\alpha ,i}({\bf r}_{i+1}^\alpha-{\bf r}_{i}^\alpha)^2/(2a^2)
  +B \sum_{\alpha ,i} ({\bf r}_i^\alpha)^2
  +D \sum_{\alpha\neq\beta ,i} d_{\alpha\beta}({\bf r}_{i}^\alpha-{\bf r}_{i}^{\beta})^2,
\end{eqnarray}
where $B$ measures the confinement to a globule and $D$, and $d_{\alpha\beta}$ are 
variational parameters specifying the vibrational freedom in a minimum and the 
replica symmetry breaking related to the configurational entropy, respectively.
We assume $d_{\alpha\beta}$ has the same structure as mean field Potts spin glasses; 
$n$ replicas are divided into $n/m$ groups, each of which has size $m$ 
and the matrix element $d_{\alpha\beta}$ is 1 if $\alpha$ and $\beta$ ($\alpha\neq\beta$) 
belong to the same group and 0 otherwise. 
It is straightforward, though cumbersome, to 
obtain $F_{\rm var}$ 
as a function of these parameters.
As in Ref.I the {\it homogeneous} glassy state 
characterized by a large constant $D$ (i.e.\ $D \gg A \gg B$)
in the reference Hamiltonian yields an asymptotic high $D$ 
expression for the free energy of the homogeneous glassy phase.
A globally appropriate homogeneous expression can be obtained by 
interpolating to the $D=0$ limit yielding the non-gradient terms in eq.(5) 
(see below).

An entropic droplet is described by an inhomogeneous situation 
where part of the polymer 
is trapped in a particular metastable state while another part can be 
in any minimum.
The exterior of the droplet has large $D$ describing 
the part trapped in a particular metastable region 
while the interior with small $D$ represents a region which can exist
in multiple states. 
For explicitness, we shall assume the interaction range $\sigma$ is small compared 
to the scale on which $D$ varies. 
Expanding the quadratic order parameter interaction 
in eq.(\ref{eq:H2}) around the midpoint of the two coordinates yields
a $(\nabla D)^2$ term describing the surface tension between mean field minima\cite{Widom}.
Defining a dimensionless Debye-Waller factor $y\equiv 2mD\sigma^2/2$, 
the free energy functional becomes in this approximation 
$F[y({\bf r})]=F_{\rm Globule}+\int {\rm d}{\bf r}
f\left(y\left({\bf r}\right)\right)$,  
where $F_{\rm Globule}$ is the constant free energy for the globule and
\protect{\begin{equation}\label{eq:Fdensity}
f\left(y\left({\bf r}\right)\right)= {m-1\over m}{3\over 2} \rho k_B T 
\ln \left[ \left(2a\over \sigma\right)^2y+1\right]
-(m-1){\beta b^2\over 4}\nu \rho^2 
\left[ \left( {y\over y+1}\right)^{3/2} 
     -{45\over 128}\sigma^2{(\nabla y)^2\over (y+1)^{7/2}}\right].
\end{equation}}
%]\narrowtext
is the inhomogeneous free energy density.
$\rho({\bf r})$ is the density of beads and 
$\nu=(\pi\sigma^2/2)^{3/2}$. 
The surface tension is proportional to $b^2$ and 
originates from the randomness of the interactions.
Although not strictly true for the Gaussian confinment model, 
we have ignored derivatives of $\rho$, which are expected to small in a 
well collapsed state with excluded volume. 
In the functional, $m$ $y$ and $\rho$ are functions of $\bf r$.
The former two play the main roles in glassy behavior. 

The uniform solutions are essentially the same as in ref.I for a contact 
interaction model. Since $\rho$ does not exhibit any peculiar behavior, 
we use the 
so-called volume approximation where the monomer density 
$\rho_\alpha({\bf r})=\rho$ is 
taken inside polymer and zero otherwise.  
Putting $y({\bf r})=\bar{y}$, we first maximize with respect to $m$ for each $\bar{y}$ 
(denoting it as $m^*(\bar{y})$).
Fig.\ref{fg:Fvsy} shows $f(m^*(\bar{y}),\bar{y})$ for several temperatures.
In the high temperature limit, there is no saddle solution and 
only  the globule state $\bar{y}=0$ is stable.
The dynamic glass transition takes place at the temperature $T_{\rm A}$ 
where a non-zero saddle solution $\bar{y}>0$ appears. 
This describes a glassy trapped state.
At temperatures below $T_{\rm A}$, there are two minima 
in $f(\bar{y})$, one for the globule ($\bar{y}=0$), which is 
thermodynamically equivalent to the sum over metastable minima,   
and the other for a particular metastable glassy minimum, ($\bar{y}=y_{\rm G}>0$).
As temperature decreases, the free energies for the two solutions approach each other and 
become degenerate at the Kauzmann temperature $T_{\rm K}$ 
where the static transition occurs. 
As in spin glasses\cite{Kurchan93}, the uniform solution yields an 
estimate of a lower bound for the barriers between 
two lowest minima as the local maximum of $f(m^*(\bar{y}),\bar{y})$, which we denote 
as occurring at $y^\ddagger$. 
Starting from zero at $T_{\rm A}$ the uniform solution barrier height is extensive and grows 
as temperature decreases 
 and saturates around $T_{\rm K}$\cite{Takada96}. 
$T_{\rm A}/T_{\rm K}$ evaluated with typical parameters for flexible polymers 
($(2a/\sigma)^2=4$, $\rho v=1$) is about $1.4$.

The globule state between $T_{\rm A}$ and $T_{\rm K}$ obtained by the ordinary 
replica theory represents 
a weighted sum of multiple local minima. 
Thus, the difference between the most probable 
free energy of a minimum and the globule free energy corresponds to 
the logarithm of the degeneracy of local minima, i.e., 
an extensive basin configurational entropy.
For the {\em short range} infinite system between $T_{\rm A}$ and $T_{\rm K}$ 
local minima are not separated by a thermodynamically large barrier but a finite one.
KW argue 
the free energy of an entropic droplet with a radius $R$ is $F\sim\Sigma R^2-Ts_{\rm c}R^3$, 
where $\Sigma$ is the surface tension and $s_{\rm c}$ is the configurational entropy density.
Since $s_{\rm c}$ disappears at $T_{\rm K}$, the critical size of the 
droplet diverges at $T_{\rm K}$ where the static glass transition occurs. 

Inhomogeneous saddle point solutions of the free energy functional 
in the replica formulation 
give a microscopic theory of the barrier analogous to the KW result for $T_{\rm A}>T>T_{\rm K}$. 
The random interactions are integrated out and so 
the interface of the replicated droplet will be spherical, 
minimizing the surface energy. 
In spherical coordinates, the stationary phase condition for $y(r)$ 
leads to an Euler-Lagrange equation\cite{Coleman85}, that
 corresponds to the Newton equation of a dissipative system, where {\sl time} 
is $r$ and a {\sl coordinate} is $y$.
For an infinite polymer, the appropriate solution has 
the boundary conditions $dy/dr|_{r=0}=0$ and $y(\infty)=y_{\rm G}$.
Since the dissipation diverges at $r=0$, the solution stays 
at $y=0$ at the beginning of trajectory ($r\sim 0$), falls after a while with an 
infinitesimal initial velocity, and stops at $y=y_{\rm G}$.
Since the dissipation monotonically decreases with $r$, there is one such trajectory
for each temperature. The critical radius $r_{\rm c}$ defined 
by the value of $r$ where $y(r_{\rm c})=y^\ddagger$ is shown as a function of 
temperature in Fig.\ref{fg:rcvsT}. Fig.\ref{fg:fvsT} shows the critical free energy 
of the droplet in an infinite system as well as that for a uniform transition 
for a finite but large $N$.
In these figures, we indicate by dashed lines the results 
obtained by the so-called {\em thin wall approximation}\cite{Coleman85}, 
which assumes a plane domain wall giving a free energy of the KW form.

For protein-like parameters,  
just below $T_{\rm A}$, 
the free energy barrier for a droplet in an infinite system is higher than 
the barrier for a 
uniform transition of a polymer with a moderate size (e.g. 1000-mer).
The interface for the droplet of small radius costs 
a considerable surface energy reflecting the difficulty of healing the interface. 
A polymer of this size escapes from traps through a nearly uniform reconfiguration.
Decreasing temperature, a localized escape mechanism begins to dominate 
(Fig.\ref{fg:fvsT}) for a sufficiently large polymer, 
but the critical radius then increases upon further cooling
until it again reaches the whole size of the polymer, because 
the entropic driving force for transition finally disappears at $T_{\rm K}$.
Thus, very close to $T_{\rm K}$, the spatially uniform transition again dominates.
The width of the spherical domain wall is $\sim 3\sigma$. This is not much smaller 
than the radius of the smaller lattice models for proteins.
The thin wall approximation is accurate for larger droplets 
(lower temperature), but it drastically 
underestimates the free energy for small droplets. 
The numerical accuracy of these results, of course, depends on how well 
the real system is approximated by the contact model but the 
trends should be robust.

We have described droplets that are local in space. 
There is an alternative escape route; an entropic droplet local 
in sequence. This can be treated using a similar reference Hamiltonian to 
eq.(\ref{eq:refH}), but with $D_i$ depending on sequence number $i$. 
Straightforward calculation shows 
that the activation barrier for the sequentially localized droplet is 
proportional to the size of system $N$ between $T_{\rm A}$ and $T_{\rm K}$,
while for droplets local in space 
the activation energy is independent of $N$ except very close to $T_{\rm K}$.
Entropic droplets local in sequence therefore do not change the story much.
Droplets local both in sequence and space correspond for glassy traps with 
the {\em foldons} of a minimally frustrated system and may be 
relevant when topological constraints are considered\cite{Panchenko95}.

The explicit droplet solutions discussed here are oversimplified. 
For structural glass and Potts glasses 
the interface has a more complex structure\cite{Kirkpatrick89}.
Wetting due to multiple states in the interface reduces the effective 
surface tension significantly. 
This leads to the Vogel-Fulcher dependence $\Delta F\sim (T-T_{\rm K})^{-1}$. 
We have not yet succeeded in quantifying this wetting phenomenon in the 
replica instanton formalism. 
Inhomogeneous states of a polymer 
(and short range Potts-type models, in general) 
may not, rigorously speaking, be described by the simple one level RSB scheme 
but require a $P(q)$ with finite width peaks.
A treatment of the interface like that for the short range SK model\cite{Parisi94}
may incorporate the wetting effect.

We are grateful to Jin Wang and John Portman for useful discussions. 
S.T. is a Postdoctoral Fellow for Research Abroad of the Japan Society for 
the Promotion of Science. P.G.W. is supported by NIH grant PHS 1 R01 GM44557.

%\setlength{\baselineskip}{0.65cm}
%************************************************************************
\narrowtext
%\twocolumn[]

%************************************************************************

%***********************************************************************
%\setlength{\baselineskip}{13pt}

\par\noindent
\vfill
%************************************************************************
\begin{figure}
\caption{Free energy density for the uniform solution as a function of 
$y=2mD\sigma^2/2$ for several temperatures. 
The value of $m$ is maximized for each $y$. Parameters used 
are $(2a/\sigma)^2=4$ and $\rho \nu=1$ and Temperatures are 
$T=0.32b$, $0.292b\sim T_{\rm A}$, $0.24b$ and $0.2039b\sim T_{\rm K}$.}
\label{fg:Fvsy}
\end{figure}

\begin{figure}
\caption{Critical radii of droplet plotted with respect to $T$. The solid (dashed) 
curve represents the spherical droplet (that by the thin approximation). 
Parameters used are the same as Fig.\protect{\ref{fg:Fvsy}}.}
\label{fg:rcvsT}
\end{figure}

\begin{figure}
\caption{Free energy barriers for droplets and uniform transitions with respect to $T$. 
Each curve is for 1)the droplet, 2)the droplet by the thin approximation, 3)the 100-mer 
uniform transition and 4)that of 1000-mer. Parameters used are the 
same as Fig.\protect{\ref{fg:Fvsy}}.}
\label{fg:fvsT}
\end{figure}

%************************************************************************
\parskip 20pt \par\noindent
\end{document}